\documentstyle[version2,aps,preprint]{revtex}
\begin{document}
\tolerance=50000
\begin{title}
A toy model of interlayer pair hopping
\end{title}
\author{V. N. Muthukumar}
\begin{instit}
Low Temperature Laboratory \\
Indian Institute of Technology, Madras 600 036, INDIA
\end{instit}
\author{G. Baskaran}
\begin{instit}
The Institute of Mathematical Sciences\\
Madras 600 113, INDIA
\end{instit}
\begin{abstract}
We extend the correlated electron model of Baskaran [Mod. Phys. Lett.
{\bf B5}, 643(1991)] to the case of coupled layers.  We show that the nature of
the non-Fermi liquid ground state leads to the absence of electron-like
quasiparticles at the Fermi surface, thereby suppressing coherent
transport of electrons between the coupled layers (``confinement'').  On
the other hand, motion of singlet pairs between the layers is not
blocked.  We also discuss how pair tunneling can be used to construct a
superconducting ground state.
\end{abstract}
The study of correlated electronic systems has attracted considerable
attention in recent times.  In particular, the arguments advanced by
Anderson and collaborators in favor of (a) failure of Fermi liquid
theory in two dimensions \cite{pwa1}, (b) spin-charge separation
\cite{pwa2}, (c) irrelevance of single particle hopping between coupled
Hubbard chains and/or planes \cite{pwa3} and (d) interlayer pair
hopping \cite{wha}
have evinced a lot of debate \cite{pp}.
Though most of these features are present
in the one dimensional Hubbard model, the two dimensional model is yet to
be solved.  Therefore, it would be instructive to examine an
interacting system of electrons which would exhibit these features in
two and/or higher dimensions.  Such a study would further our
understanding of strongly correlated electronic systems.

A step in this direction was taken by one of us \cite{gb}
who proposed
and solved a model of strongly interacting fermions in d-dimensions.
The advantage of this model is that it can be solved exactly to show the
breakdown of Fermi liquid theory and the emergence of spin-charge
decoupling, thus illustrating two of the four features we listed above.  In
this letter, we extend this work to the case of coupled layers and
examine the remaining two features, viz., single particle hopping
between the coupled layers and interlayer pair hopping.  We show that
while the non-Fermi liquid ground state leads to a suppression of
coherent one-electron motion between the coupled layers, motion of
singlet pairs is not blocked.  Both these results are obtained exactly.
Finally, we discuss how long range superconducting order can be
obtained in the presence of a residual intralayer pairing
mechanism acting in conjunction with pair tunneling and how the scale of
the superconducting transition temperature $T_c$ is governed by the
interlayer hopping matrix element $t_{\perp}$.

We begin by writing the model hamiltonian of ref.\cite{gb},
\begin{equation}
H_o \ = \ \sum_{k\sigma} \epsilon_k \  c^{\dagger}_{k\sigma}  \
c_{k\sigma} \ +
\ J~{\sum_{kk^\prime}}^\prime \vec{S}_k . \vec{S}_{k^\prime} \ ,
\end{equation}
where $\vec{S}_k = c^{\dagger}_{k\alpha} \ \vec{\sigma}_{\alpha\beta}
\ c_{k\beta}$ is the spin operator ($ S = {1 \over 2}$) in momentum space
and $\epsilon_k$ is assumed to be of the form
$\displaystyle{{{\hbar}^2 k^2 \over 2m}}$
for simplicity \cite{koh}.  The primed sum indicates the absence of the
$k=k^\prime$ term in the summand.  Thus $H_o$ is a $t-J$ model in
momentum space.  As shown in [6], $H_o$ can be solved exactly in
d-dimensions and the ground state $\vert G \rangle_o$ is given by
\begin{equation}
\vert G \rangle_o \ = \  P_S \ \prod_{k_1 < \vert k \vert
< k_2} \ c^{\dagger}_{k\sigma_k} \ \prod_{\vert k \vert < k_1}
\ c^{\dagger}_{k\uparrow}\  c^{\dagger}_{k\downarrow} \ \vert 0 \rangle \ ,
\end{equation}
where $ P_S$ is a singlet projection operator and
$\{\sigma_k\}$ is any
$k$-dependent spin configuration in momentum space in the singly
occupied annulus of radii $k_1$ and $k_2$ such that the condition \\
$\sum_{k_1 < \vert k \vert < k_2} S_k^z$ = $0$ is satisfied.
The radii $k_1$ and $k_2$ are determined by the equations
\begin{eqnarray*}
\epsilon_{k_2} - \epsilon_{k_1} & = & J  \\
\sum_{k_1 < \vert k \vert < k_F} & = &
\sum_{k_2 > \vert k \vert > k_F} \ \ ,
\end{eqnarray*}
where $k_F$ is the Fermi wave vector of the noninteracting Fermi sea.
The ground state is therefore a filled doubly occupied Fermi sea upto
radius $k_1$ ; between radii $k_1$ and $k_2$, it is singly occupied and
the electrons inside this annulus form a total singlet in momentum
space.  Therefore, the momentum distribution function $n_k$ (at zero
temperature) is given by
$$
n_k = \left\{ \begin{array}{ll}
       1 & \mbox {if $ \vert k \vert < k_1$}\\
       {1 \over 2} & \mbox {if $k_1 < \vert k \vert < k_2$}
       \end{array}
       		\right.
$$
As there is no discontinuity in the occupancy at $k_F$, viz.,
$Z_{k_F}$ = $0$, the result implies a complete breakdown of
Fermi liquid theory.  (We note  here that it is possible to restrict the
range of the $k$-space superexchange interaction to a shell of desired
width about the Fermi surface, leading to an asymptotic failure of Fermi
liquid theory as opposed to a complete breakdown as above.)
Since the
ground state is known exactly, this feature can also be seen by writing
down the explicit form of the one-electron Green's function
$G(k,\omega)$.  $G(k,\omega)$ is defined in the usual manner as
$$
G(k,\omega)\ =\ -i\ \int_{-\infty}^{+\infty}\ {\rm e}^{-i\omega t}
\ \langle \ Tc_{k\uparrow}(t)~c_{k\uparrow}^{\dagger}(0)\ \rangle\ dt\ \ .
$$
We use the equation of motion method to evaluate $G(k,\omega)$
exactly.  The time dependence of $c_{k\uparrow}$ will be
governed by commutators that occur in the equation of motion for
$G(k,\omega)$ such as,
$$
\left[~H,~c_{k\uparrow}~\right]\ =\ -(\epsilon_k - {3J \over 4})
{}~c_{k\uparrow} - J~c_{k\downarrow}~\sum_p~S^-_p\ -\
J~c_{k\uparrow}~\sum_p~S^z_p\ \ .
$$
The expectation values of such commutators are evaluated with
respect to one of the infinitely many ground states.
Since the ground state is a total singlet it follows that
$$
\langle c^{\dagger}_{k\uparrow}~\left[~H,~c_{k\uparrow}~\right] \rangle
\ =\ -(\epsilon_k - {3J \over 4})~\langle c^{\dagger}_{k\uparrow}c_{k\uparrow}
 \rangle
$$
(Here, we note that it is not necessary for us to know the exact form of
the ground state wave function.  We only need to know that the ground
state is a total singlet.)  A similar line of reasoning will hold good
for the higher order terms in the expression for $c_{k\uparrow}(t)$.
This simplifies our calculation of $G(k,\omega)$ and we find that
$G(k,\omega)$ is given by
$$
G(k,\omega) \ = \ \displaystyle{{1 \over i\omega - (\epsilon_k - {3J
\over 4})}}  .
$$
The pole that is present at $\omega~=~\epsilon_k$ in the case of
the non interacting system is absent and instead we now obtain a
pole at $\omega~=~\epsilon_k~-~{3J \over 4}$.  The new pole has
its residue contributed by one of the many ($\sim N$) degenerate
states.  All these states are characterized by a double occupancy
at point $k$ and one unpaired spin elsewhere in the singly
occupied $k$-space annulus of radii $k_1$ and $k_2$.  It is this
which gives a potentially incoherent character to the spectral
weight.  For example, by choosing $J$ in (1) to be dependent on
momentum, it is possible to lift the degeneracy of these states
and this would spread the pole to a truly incoherent background.
This behaviour should be contrasted with the shifted pole
structure in say, a mean field CDW or BCS state where the new
poles essentially correspond to one or two new eigen states that
are not incoherent.  Thus it is clear that the shifting of the pole at
$k_F$ is not a simple energy denominator effect or an energy
shift effect. We can also see this by considering the motion of
singlets, viz., the Green's function of a singlet pair
${\cal S}(k,k^\prime;\omega)$.
We define
$$
{\cal S} (k,k^\prime ;\omega) \ = \ -i\int_{-\infty}^{+\infty} \
\langle \ T\ b_{kk^\prime} (t)\  b^\dagger_{kk^\prime} (0)\  \rangle
\ e^{-i\omega t}\  dt  \ \ ,
$$
where
$$
b_{kk^\prime}^{\dag} \ = \ {1\over\sqrt{2}} \ (c^{\dag}_{k\uparrow}\
c^{\dag}_{k^\prime \downarrow} - c^{\dag}_{k\downarrow} \ c^{\dag}_{k^\prime
\uparrow})  \ ,
$$
is the creation operator of a singlet bond between $k$ and $k^\prime$.
As in the case of $G(k,\omega)$, we can evaluate
${\cal S} (k,k^\prime;\omega)$ exactly.  We find that
$$
{\cal S}(k,k^\prime;\omega) \ = \
\displaystyle{{1 \over i\omega - (\epsilon_k + \epsilon_{k^\prime})}}
\ \ ,
$$
i.e., the pole gets shifted back and the two particle (singlet)
Green's function has the same form as that of the non
interacting system!
This is essentially due to the
singlet correlations induced by the $k$-space superexchange term in
$H_o$.  Thus we see that while the pole at $k_F$ disappears in the case
of the one-electron Green's function it persists in the two
electron (singlet) Green's function.
Obviously, this  cannot happen if the vanishing of the pole
were a simple energy denominator or an energy shift effect.

We now ask if coherent hopping of electrons between coupled layers is
possible in the presence of an interlayer hopping term
$$
\sum_{k\sigma} t_{\perp}(k)\  (\ c^{\dag}_{k\sigma}~d_{k\sigma} +
h.c.\ ) \ \ ,
$$
Here, $t_{\perp}(k)$ is the dispersion along the c-axis
and $c$ and $d$ represent the coupled layers.  (Though our results are valid
for d-dimensions, we shall continue to use the term ``layer''
throughout.)  The hamiltonian $H$ is now given by
$$
\begin{array}{ccll}
H& =& \sum_{k\sigma} \epsilon_k\  (c^{\dag}_{k\sigma} c_{k\sigma} +
d^{\dag}_{k\sigma} d_{k\sigma}) & + J~{\sum^\prime}_{kk^\prime}
\ \vec{S}^c_k . \vec{S}^c_{k^\prime} +\vec{S}^d_k . \vec{S}^d_{k^\prime} \\
 & & &+ \sum_{k\sigma} t_{\perp}(k)\  (c^{\dag}_{k\sigma} d_{k\sigma}
 + {\rm h.c.} ) .
\end{array}
$$
We check for c-axis current by evaluating the Green's function
$$
G^\perp (k,\omega)\ =\ -i \int_{-\infty}^{+\infty}\ \langle\  T
c^{\dag}_{k\sigma}(t) d_{k\sigma}(0) \ \rangle e^{-i\omega t} \ dt  \ .
$$
If this has a pole structure, it would mean that electrons can be transported
along the c-direction.  In such a case, the interlayer coupling would be
termed a ``relevant operator''.  We evaluate $G^{\perp} (k,\omega)$ by
defining the operators $\alpha_{k\sigma}$ and $\beta_{k\sigma}$ ,
\begin{eqnarray*}
\alpha_{k\sigma} & = & {1 \over \sqrt{2}} \ (c_{k\sigma} + d_{k\sigma})\\
\beta_{k\sigma} & = & {1 \over \sqrt{2}} \ (c_{k\sigma} - d_{k\sigma})\ ,
\end{eqnarray*}
and reexpressing the hamiltonian in terms of these operators.  It is
easily verified that $H$ can be rewritten as
\begin{eqnarray*}
H = & \sum_{k\sigma} \left[\  (\epsilon_k - t_{\perp}(k))\
\beta^{\dag}_{k\sigma} \beta_{k\sigma}
+ (\epsilon_k + t_{\perp}(k)) \ \alpha^{\dag}_{k\sigma} \alpha_{k\sigma}
\right] \\
& +\displaystyle{{J \over 2}}~{\sum_{kk^\prime}}^\prime \vec{S}^\alpha_k
.\vec{S}^\alpha_{k^\prime} + \vec{S}^\beta_k . \vec{S}^\beta_{k^\prime}
+ 2 \vec{S}^\alpha_k . \vec{S}^\beta_{k^\prime} \\
& + \displaystyle{{J \over 2}}~{\sum_{kk^\prime}}^\prime \
( \vec{S}^{\alpha \beta}_k + \vec{S}^{\beta \alpha}_k)
. ( \vec{S}^{\alpha \beta}_{k^\prime} + \vec{S}^{\beta \alpha}_{k^\prime})
\ ,
\end{eqnarray*}
where we have defined
\begin{eqnarray*}
\vec{S}^{\alpha}_k & = & {1 \over 2}\  \alpha^{\dag}_{k\mu}
\ \vec{\sigma}_{\mu \nu} \ \alpha_{k\nu}\ , \\
\vec{S}^{\alpha \beta}_k & = & {1 \over 2}\  \alpha^{\dag}_{k\mu}
\ \vec{\sigma}_{\mu \nu}\  \beta_{k\nu}\ , \ {\rm etc.}
\end{eqnarray*}
Since $\langle \ T c^\dagger_{k\sigma}(t)\  d_{k\sigma}(0)\  \rangle$ =
$\langle\  T d^\dagger_{k\sigma}(t)\  c_{k\sigma}(0)\  \rangle$ by symmetry, we
can write
$$
G^{\perp} (k,\omega) \ = \ -i\int^{+\infty}_{-\infty} e^{-i\omega t}
\left[\  \langle\  T \alpha^{\dag}_{k\sigma}(t)\  \alpha_{k\sigma}(0)
\ \rangle -
\langle\  T \beta^{\dag}_{k\sigma}(t)\  \beta_{k\sigma}(0)\
\rangle\  \right]\ .
$$
Though we could not write down the exact ground state wave
function we have nevertheless been able to obtain the exact form of
$G^{\perp}(k,\omega)$ by exploiting the symmetry between the two layers $c$
and $d$ and by using the fact that the ground state
wave function $\vert GS \rangle$ must be a total singlet, viz.,
$$
\sum_k \  (S^{c-}_k + S^{d-}_k)\  \vert GS \rangle \equiv 0 \
,
$$
(where $S^{c-}$ and $S^{d-}$ are the usual spin lowering operators) as
well as the symmetry between the coupled layers by virtue of which
$$
\sum_k S^{cz}_k - S^{dz}_k \vert GS \rangle \ \equiv 0
\ \ {\rm etc.,}
$$
In terms of the $\alpha$ and the $\beta$ operators, we then have the
following conditions satisfied by the ground state $\vert GS \rangle$.
\begin{eqnarray*}
\sum_k \ ( S^{\alpha -}_k + S^{\beta -}_k ) \ \vert GS \rangle & = & 0\
,\\
\sum_k \ ( S^{\alpha z}_k + S^{\beta z}_k ) \ \vert GS \rangle & = & 0
\end{eqnarray*}
and
\begin{eqnarray*}
\sum_k\ ( S^{\alpha \beta -}_k + S^{\beta \alpha -}_k )\ \vert GS \rangle
& = & 0\ ,\\
\sum_k\ ( S^{\alpha \beta z}_k + S^{\beta \alpha z}_k )\ \vert GS \rangle
& = & 0.
\end{eqnarray*}
Using these conditions, we arrive at the form of the Green's function
for interlayer motion,
$$
G^\perp (k,\omega) \propto \
\left[\  \displaystyle{{\langle \alpha_{k\sigma}^{\dag} \alpha_{k\sigma}
\rangle
\over i\omega - (\epsilon_k - t_{\perp}(k) - {3J \over
4})}}
+ \displaystyle{{\langle \beta_{k\sigma} \beta^{\dag}_{k\sigma} \rangle
\over i\omega - (\epsilon_k + t_{\perp}(k) - {3J \over
4})}} \ \right] \ .
$$
Since in general, $\langle \alpha^{\dag}_{k\sigma} \alpha_{k\sigma}
\rangle$ and
$\langle \beta^{\dag}_{k\sigma} \beta_{k\sigma} \rangle$ $\leq$ 1, we
see again that there will be no pole structure in the Green's function
for energies less than ${3J \over 4}$ as in the single layer
case. As discussed before, these two poles are also potentially
incoherent and modifying the form of $J$ would spread the pole
to an incoherent background.
Thus the non Fermi liquid nature of the ground state persists in the presence
of interlayer hopping and this causes the bare hopping matrix element to
get renormalized to zero.  This is the phenomenon of ``confinement'' which
is being studied extensively in coupled Hubbard chains/planes.

We now turn our attention to the Green's function for singlet transport
between the layers ${\cal S}^{cd}(k,k^{\prime};\omega)$.
It is not difficult to see that ${\cal S}^{cd}(k,k^{\prime};\omega)$ is given
by
$$
{\cal S}^{cd}(k,k^{\prime};\omega) \ \propto
\left[ \displaystyle{{\langle \alpha_{k\uparrow}^{\dag} \alpha_{k\uparrow}
\rangle\ \langle \alpha_{k^\prime \downarrow}^{\dag}
\alpha_{k^\prime \downarrow} \rangle
\over i\omega - (\epsilon_k - t_{\perp}(k) + \epsilon_{k^\prime} -
t_{\perp}(k^\prime))}}
+ \displaystyle{{\langle \beta_{k\uparrow}\beta_{k\uparrow}^{\dag}  \rangle
\ \langle \beta_{k^\prime \downarrow}\beta_{k^{\prime}\downarrow}^{\dag}
 \rangle
\over i\omega - (\epsilon_k + t_{\perp}(k) + \epsilon_{k^\prime} +
t_{\perp}(k^\prime))}} \right] \ .
$$
The pole structure is got back when singlet transport is considered.
So we see that while one electron motion between the coupled layers is
blocked (the corresponding Green's function being completely
incoherent), singlets can be transported freely.
Another way of interpreting this result is to say that while the state
describing an electron-hole excitation between the layers,
$c^{\dag}_{k\sigma} d_{k\sigma} \vert N^c;N^d \rangle$ suffers an
orthogonality catastrophe
$$
\langle (N+1)^c;(N-1)^d\ \vert \ c^{\dag}_{k\sigma} d_{k\sigma} \ \vert
N^c;N^d \rangle \ =\ 0 ,
$$
there is a finite overlap between states that are connected by
pair hopping, viz.,
$$
\langle (N-2)^c;(N+2)^d\ \vert \ b^{d\dag}_{kk^\prime}
\ b^c_{kk^\prime} \ \vert
N^c;N^d \rangle \ \ \neq 0 .
$$

Now that we have shown that it is possible to transport pairs of electrons
between the two coupled layers, the next obvious question would be to
ask if such a process leads to superconductivity.  Since we have
calculated the interlayer two particle Green's function, we
can,in principle calculate the Josephson current between the
coupled layers and discuss the nature of the superconducting
state.  But we adopt the following
perturbative approach which not only illustrates
the pair hopping mechanism but  also enables us to compare
our results with those obtained earlier in the context of high
temperature superconductivity from interlayer tunneling \cite{wha},
\cite{phen}, \cite{chak}.
As we have shown
that the interlayer single electron Green's function does not have a
pole  at $\epsilon_k$
whereas the two electron (singlet) Green's function does , it is
enough to consider only pair hopping processes that are generated
perturbatively.
Let us consider the interlayer hopping as perturbing the initial ground
state $\vert G \rangle_o$ by breaking, for instance, a singlet bond
between momentum states $k$ and $-k$ with the rest of the wave function
being undisturbed.  This results in the creation of a particle hole
excitation between the layers which we treat as an intermediate state in
perturbation theory.  It is easy to see that  second order perturbation
theory leads to an effective interaction of the form
\begin{equation}
- \displaystyle{{1 \over {3J \over 4}}}\
\sum_k t_{\perp}^2(k) \
c^{\dag}_{k\uparrow} c^{\dag}_{-k \downarrow} d_{-k\downarrow}
d_{k\uparrow} \ .
\end{equation}
This hamiltonian resembles the holon pair hopping hamiltonian that was
first introduced by Wheatley, Hsu and Anderson \cite{wha}.  However
there is an important difference since the pair hopping interaction
above is extremely local in $k$-space.  This is because the interlayer
hopping matrix element $t_{\perp}(k)$ (which generates the pair hopping
process) conserves the momentum parallel to the layers.  The
significance of this fact and its implications were first discussed by
Anderson \cite{onek}. Such a pair hopping hamiltonian has been
studied recently by several people \cite{phen},  \cite{chak}.  In
particular, one can for instance do a mean field calculation with the
pairing interaction as given by (3) and obtain the superconducting
transition temperature
$$
k_B T_c~\sim ~ {\displaystyle {t_\perp^2(k)_{max} \over J}}~~.
$$
Note that $T_c$ depends on the pairing interaction in a simple manner
unlike the BCS case.  But the extreme ($k$-space) locality of the
pairing interaction will suppress any finite temperature phase
transition since the phases of various pairing amplitudes $\langle
c^{\dag}_{k\uparrow}c^{\dag}_{-k\downarrow}\rangle$ do not get
correlated. However, it is clear that any residual pairing interaction
such as an intralayer BCS interaction which is non local in $k$-space
will correlate the phases of various pairing amplitudes and will result
in long range superconducting order.  It should be emphasized that the
scale of $T_c$ will still be set by the strength of the pair
tunneling interaction  rather than that of the intralayer interaction
as long as ${t_{\perp}^2 \over J}$~ $>$~$V_{BCS}$ \cite{chak}.

Before we conclude we would like to point out that the model under
consideration was
introduced to mimic the single occupancy constraint in a realistic model
such as the large $U$ Hubbard model, where it is speculated that forward
scattering of a pair of electrons becomes singular asymptotically close
to the Fermi surface.  The model discussed in this letter does not show
this asymptotic behaviour but leads to a complete breakdown of Fermi
liquid theory.  We believe this model can be made more realistic by
introducing a suitable momentum dependent $J$ in (1).  This is a topic
of future study.

To summarize, we have been able to illustrate several features of
interlayer pair hopping that have been proposed in connection with high
temperature superconductivity such as ``confinement'', $k$-space locality of
the pair hopping interaction and the analytic dependence of $T_c$ on the
strength of the pair hopping interaction.  Our results assume added
significance owing to the fact we have derived these features from a
microscopic hamiltonian, albeit a toy one.
\section{Acknowledgements}
One of us (V.N.M.) acknowledges financial support from the Program
Management Board for Superconductivity, Department of Science and
Technology, India.

\end{document}